\documentclass[aps,prc,twocolumn,preprintnumbers,amsmath,amssymb,superscriptaddress,footnote]{revtex4-2}
\usepackage{epsfig,bm}
\usepackage{ulem}
\usepackage{amsmath,color}

\begin{document}

\title{Three-body description of $^9$C: Role of low-lying resonances in breakup reactions}

\author{Jagjit Singh}
\email[]{jsingh@rcnp.osaka-u.ac.jp}
\affiliation{Research Center for Nuclear Physics (RCNP), Osaka University, Ibaraki 567-0047, Japan}

\author{Takuma Matsumoto}
\email[]{matsumoto@phys.kyushu-u.ac.jp}
\affiliation{Department of Physics, Kyushu University, Fukuoka 819-0395, Japan}

\author{Tokuro Fukui}
\email[]{tokuro.fukui@riken.jp}
\altaffiliation[Present address: ]{RIKEN Nishina Center, Wako 351-0198, Japan}
\affiliation{Yukawa Institute for Theoretical Physics, Kyoto University, Kitashirakawa Oiwake-Cho, Kyoto 606-8502, Japan}

\author{Kazuyuki Ogata}
\email[]{kazuyuki@rcnp.osaka-u.ac.jp}
\affiliation{Research Center for Nuclear Physics (RCNP), Osaka University, Ibaraki 567-0047, Japan}
\affiliation{Department of Physics, Osaka City University, Osaka 558-8585, Japan}
\affiliation{Nambu Yoichiro Institute of Theoretical and Experimental Physics (NITEP), Osaka City University,
Osaka 558-8585, Japan}
\date{\today}

\begin{abstract}
 \edef\oldrightskip{\the\rightskip}
 \begin{description}
  \rightskip\oldrightskip\relax
  \item[Background]
    The $^9$C nucleus and related capture reaction, ${^8\mathrm{B}}(p,\gamma){^9\mathrm{C}}$, have been intensively studied with an astrophysical interest. Due to the weakly-bound nature of $^9$C, its structure is likely to be described as the three-body (${^7\mathrm{Be}}+p+p$). Its continuum structure is also important to describe reaction processes of $^9$C, with which the reaction rate of 
    the ${^8\mathrm{B}}(p,\gamma){^9\mathrm{C}}$ process have been extracted indirectly.
  \item[Purpose]
    We preform three-body calculations on $^9$C and discuss properties of its ground and low-lying states via breakup reactions.
  \item[Methods]
    We employ the three-body model of $^9$C using the Gaussian-expansion method combined with the complex-scaling method. This model is implemented in the four-body version of the continuum-discretized coupled-channels method, by which breakup reactions of $^9$C are studied.
    The intrinsic spin of $^7$Be is disregarded.
  \item[Results] 
    By tuning a three-body interaction in the Hamiltonian of $^9$C, we obtain the low-lying $2^+$ state with the resonant energy $0.781$\,MeV and the decay width $0.137$\,MeV, which is consistent with the available experimental information and a relatively high-lying second $2^+$ wider resonant state. Our calculation predicts also sole $0^+$ and three $1^-$ resonant states. We discuss the role of these resonances in the elastic breakup cross section of $^9$C on $^{208}$Pb at $65$ and $160$\,MeV/nucleon.
  \item[Conclusions] 
    The low-lying $2^+$ state is probed as a sharp peak of the breakup cross section, while the $1^-$ states enhance the cross section around $3$\,MeV. 
    Our calculations will further support the future and ongoing experimental campaigns for extracting astrophysical information and evaluating the two-proton removal cross-sections.
\end{description}
\end{abstract}

\maketitle

\section{Introduction}

In stellar nucleosynthesis, proton capture reaction of $^8$B, 
${^8\mathrm{B}}(p,\gamma){^9\mathrm{C}}$, is expected to drive an explosive hydrogen burning 
($pp$ chain) and serves as a possibly alternative path to the synthesis 
of the CNO elements~\cite{Wiescher89}. Due to the difficulties in measuring 
the ${^8\mathrm{B}}(p,\gamma){^9\mathrm{C}}$ cross section at very low energies, 
several experiments of alternative reactions have been performed to indirectly 
determine the astrophysical $S$-factor~\cite{Beaumel01,Trache02,Moto03}. 
The significant discrepancy in the previous measurements for the determined astrophysical $S$-factor~\cite{Fukui12}, has triggered off the new experimental study on the $^9$C breakup~\cite{Chilug19,Chilug20}. 
For studying $^9$C-breakup reactions as well as ${^8\mathrm{B}}(p,\gamma){^9\mathrm{C}}$, the structure of $^9$C plays a key role.

On the experimental side, the ground and low-lying states of $^9$C were measured via multi-nucleon transfer reaction~\cite{Cerny64,Ben74,Golovkov91}. 
Recently the higher excited states were explored in Refs.~\cite{Rogachev07,Brown17,Hooker19}.
It is noteworthy that the first positive parity $5/2^+$ state was identified~\cite{Hooker19} in the mass number 9 and isospin $3/2$ systems. Elastic scattering angular distributions of $^9$C on $^{208}$Pb 
target at $25.2$\,MeV/nucleon has been measured recently \cite{Yang18}. 
On the theoretical side, in Ref.~\cite{Arai01}, the four-body ($\alpha+{^3\mathrm{He}}+p+p$) calculation was performed to study the ground-state property of $^9$C, 
while its excited states were predicted by the antisymmetrized molecular dynamics~\cite{Kanada01} and the continuum shell model~\cite{Volya14}.  
Furthermore, the astrophysical $S$-factor of ${^8\mathrm{B}}(p,\gamma){^9\mathrm{C}}$ was estimated theoretically, with assuming $^9$C of the ${^8\mathrm{B}}+p$ configuration, 
by the microscopic cluster model~\cite{DESCOUVEMONT99} 
and the continuum-discretized coupled-channels method (CDCC)~\cite{Kamimura86,Austern87,Yah12}
applied to the $^9$C breakup processes~\cite{Fukui12} 
and the transfer reaction ${^8\mathrm{B}}(d,n){^9\mathrm{C}}$~\cite{Fukui15}.  

We remark that $^9$C is only bound by $1.436$\,MeV \cite{Wang17} with respect to ${^7\mathrm{Be}}+p+p$ threshold, and all of its excited states are in the continuum; there exists a ${^8\mathrm{B}}+p$ threshold 
just below the three-body threshold by $0.137$\,MeV. 
It is important to accurately reproduce these threshold energies. The  ${^7\mathrm{Be}}+p+p$ three-body model is suitable for this purpose as explained below. Furthermore, the result of the three-body model calculation can be implemented in CDCC and one can thereby describe dynamical processes of $^9$C. 
This will be essential to extract the astrophysical $S$-factor from experimental data 
in future. 

In past, the $^{9}$C along with other light mirror nuclei was studied within a three-body model 
based on hyperspherical framework \cite{Timofeyuk08,Timofeyuk208}. 
The prime focus of this work was to study the mirror spectroscopic factors, 
mirror asymptotic normalization coefficients (ANCs), and pre-asymptotic abnormalities \cite{Timofeyuk03} 
in the overlap functions for the various three-body mirror systems.
These are interesting subjects but beyond the scope of the present study. 
The purpose of this study is to describe the ground and continuum states of $^9$C by 
a three-body model and show how breakup reaction observables reflect these structures for the first time. 
To achieve this, the three-body ground state and discretized-continuum states of
$^9$C are obtained with the Gaussian-expansion method (GEM)~\cite{Hiyama03}.
For searching the resonances in the low-lying continuum states, we make 
use of the complex-scaling method (CSM)~\cite{Agu71,Bal71,Aoy06,Mat10}. 
We then discuss the role of these low-lying resonances in the elastic breakup cross sections of $^9$C on the $^{208}$Pb target at $65$ and $160$\,MeV/nucleon. 
The ${^9\mathrm{C}}+{^{208}\mathrm{Pb}}$ scattering is described by the CDCC
based on the four-body (${^7\mathrm{Be}}+p+p+{^{208}\mathrm{Pb}}$) model. 
Although the four-body CDCC is very well tested for reactions involving two-neutron halo systems~\cite{Mat04,Mat06,Mat19,Ogawa2020}, it is the first time we are extending its implementation for the proton-rich nucleus.

This paper is organized as follows. 
In Sec. II, we briefly describe the theoretical framework used for the present study. In Sec. III, we show our main results for the low-lying continuum spectrum of $^9$C and its breakup reaction. Finally, we give the conclusions in Sec. IV.

\section{Theoretical Framework}
\begin{figure}[htbp]
 \includegraphics[scale=0.25]{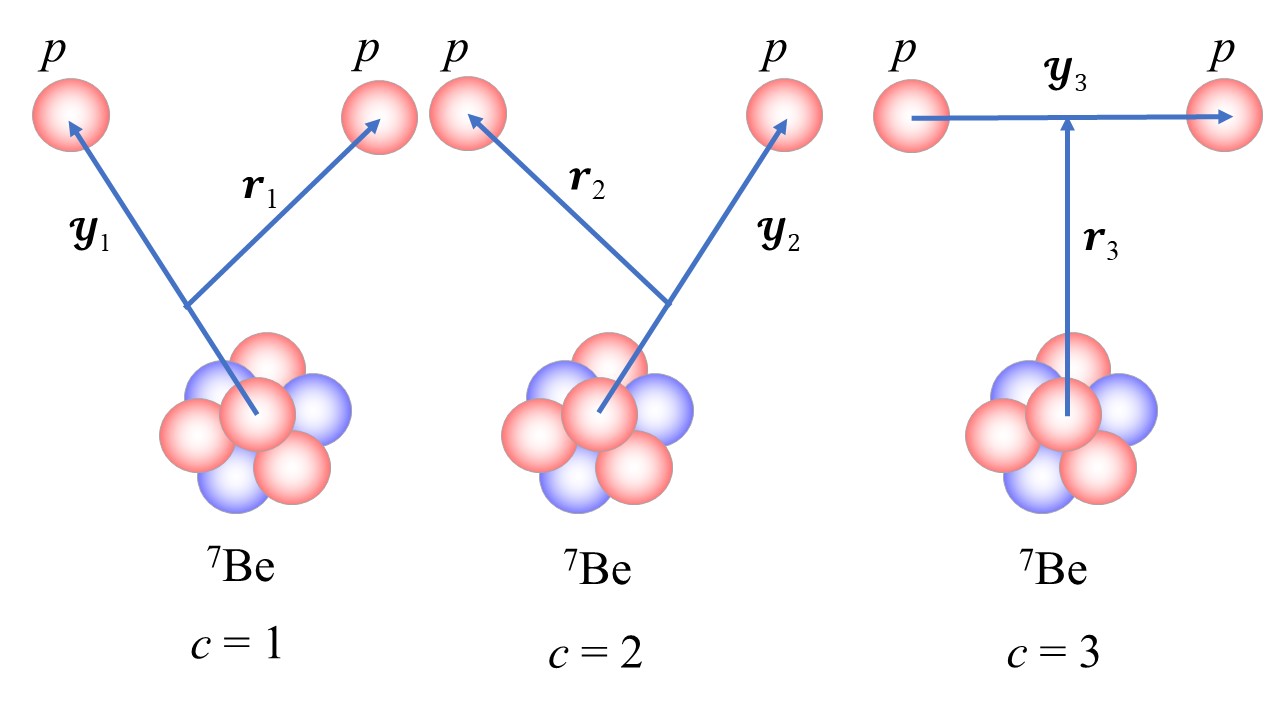}
 \caption{
 Three sets of the Jacobi coordinates $\bm{\xi}_c$~$=$~$(\bm{r}_c,~\bm{y}_c)$ in the ${^7\mathrm{Be}}+p+p$ three-body model. Each set is identified with $c$ ($c~=~1,~2,~3$).}
 \label{fig1}
\end{figure}

We describe $^{9}$C by the ${^7\mathrm{Be}}+p+p$ three-body model. 
The ground and discretized-continuum states of $^{9}$C are built within the framework of the GEM by using Gaussian basis functions~\cite{Hiyama03}.
The three-body wave function $\Phi_{I^\pi \nu}(\bm{\xi}_c)$ of $^{9}$C is characterized by the spin $I$, parity $\pi$, eigenenergy index $\nu$, and 
$\bm{\xi}_c$~$=$~$(\bm{r}_c,~\bm{y}_c)$ with $c~=~1,~2,~3$ being three sets of the Jacobi coordinates (shown in Fig.~\ref{fig1}), and it is obtained by diagonalizing the three-body Hamiltonian:
\begin{eqnarray}
 h&=&K_{r}+K_{y}+V_{cp}+V_{cp}+V_{pp}+V_{cpp}.
 \label{ham}
\end{eqnarray}
Here, $K_{r}$ and $K_y$ are the kinetic energy operators associated with the Jacobi coordinates $\bm{r}_c$ and $\bm{y}_c$ shown in Fig.~\ref{fig1},
while $V_{cp}$ ($V_{pp}$) is a two-body interaction between the $^{7}$Be and a proton (two protons), and 
$V_{cpp}$ is a phenomenological three-body force (3BF).
The antisymmetrization of $\Phi_{I^\pi \nu}$ with respect to the exchange between the two valence protons is explicitly taken into account, whereas the exchange between each valence proton and a nucleon in $^7$Be is approximately treated by the orthogonality condition model~\cite{Sai69}.

For detailed insights of the three-body continuum of $^{9}$C, we employ the CSM, 
in which the radial part of each Jacobi coordinate is transformed as $\bm{r}_c\to \bm{r}_ce^{i\theta}$ and $\bm{y}_c\to \bm{y}_ce^{i\theta}$ with the scaling angle $\theta$, and thus, $h$ is rewritten as $h^{\theta}$ accordingly. 
The diagonalization of  $h^{\theta}$ results in the eigenstates $\varphi_{\gamma}^{\theta}$ with complex eigenenergies $\varepsilon_{\gamma}^{\theta}$, where $\gamma$ is a collective index; $\gamma=(I^\pi,\nu)$. 
In the CSM, a resonance is identified as an eigenstate on the complex-energy plane isolated from other nonresonant states; 
the real and imaginary parts of the eigenenergy represent the resonant energy and a half of the decay width, respectively. 
In the present analysis, we take $I^{\pi}=0^{+}$, $1^{-}$, and $2^{+}$ states of $^9$C, 
and $\varphi_{\gamma}^{\theta}$ are also used in calculating a
continuous breakup energy spectrum from discrete breakup cross sections
obtained by CDCC, as discussed below.

The total wave function $\Psi$ of the ${^9\mathrm{C}}+{^{208}\mathrm{Pb}}$ reaction system is obtained by solving the Schr\"{o}dinger equation,
\begin{eqnarray}
\left(
K_R +h + V_{pT}+V_{pT}+V_{cT}
-E\right)\Psi(\bm{\xi},\bm{R})=0,
 \label{Sch}
\end{eqnarray}
where $\bm{R}$ and $\bm{\xi}$ represent the coordinates between $^{208}$Pb and
the center-of-mass (c.m.) of $^9$C and the intrinsic coordinate of $^9$C,
respectively.
The kinetic energy operator $K_{R}$ is associated with $\bm{R}$,
while $V_{pT}$ and $V_{cT}$ are the potentials consisting of the nuclear and Coulomb parts 
of the $p$-$^{208}$Pb and $^7$Be-$^{208}$Pb systems, respectively. 
Thus, the nuclear and Coulomb breakup processes are simultaneously taken into account.

In CDCC, we assume that the scattering takes place in a model space spanned by
\begin{eqnarray}
 {\cal P}&=&\sum_{\gamma}|\Phi_{\gamma}\rangle\langle\Phi_{\gamma}|.
\end{eqnarray}
The total wave function $\Psi(\bm{\xi},\bm{R})$ is then approximated into
\begin{eqnarray}
 \Psi(\bm{\xi},\bm{R}) \approx
  {\cal P}\Psi(\bm{\xi},\bm{R})&=&
  \sum_{\gamma}\chi_\gamma(\bm{R})
  \Phi_{\gamma}\phi_{\rm c}\nonumber\\
  &\equiv& \Psi_{\mathrm{CDCC}}(\bm{\xi},\bm{R}), \label{Eq:Psi}
\end{eqnarray}
where $\chi_\gamma$ is the relative wave
function regarding $\bm{R}$, and $\phi_{\rm c}$ is an internal wave
function of $^7$Be.
Inserting Eq.~\eqref{Eq:Psi} into Eq.~\eqref{Sch} leads to a set of
coupled equations for $\chi_\gamma$:
\begin{align}
\left[
 K_R+U_{\gamma\gamma}(\bm{R})-(E-\varepsilon_\gamma)
 \right]&\chi_\gamma(\bm{R})
 \nonumber\\
 =-\sum_{\gamma'\ne\gamma}&U_{\gamma\gamma'}(\bm{R})
 \chi_{\gamma'}(\bm{R})
 \label{Eq:cc}
\end{align}
with $\varepsilon_\gamma=\langle\Phi_\gamma|h|\Phi_\gamma\rangle$.
This is called the CDCC equations, which are solved under the standard boundary condition~\cite{Yah12}.
In the microscopic four-body CDCC, coupling potentials $U_{\gamma\gamma'}$ between $\Phi_\gamma$ and $\Phi_{\gamma'}$ are obtained by the folding procedure as 
\begin{eqnarray}
 U_{\gamma\gamma'}(\bm{R})=
 \langle\Phi_\gamma|\left[V_{pT}+V_{pT}+V_{cT}\right]|\Phi_\gamma'\rangle.
   \label{rhogg}
\end{eqnarray}

By solving Eq.~(\ref{Eq:cc}), one obtains the transition matrix element $T_{\gamma}$~\cite{Mat10}, from which the cross sections to the ground and discretized-continuum states of $^{9}$C can be evaluated.
To obtain a continuous breakup energy spectrum, we employ the
smoothing method based on the CSM proposed in Ref.~\cite{Mat10}.
Consequently, the double differential breakup cross section reads
\begin{eqnarray}
\frac{d^2 \sigma}{d\varepsilon d\Omega}
=
\frac{1}{\pi}
{\rm Im}
\sum_{\gamma'}
\frac{T_{\gamma'}^{\theta} \tilde{T}_{\gamma'}^{\theta} }
{\varepsilon-\varepsilon_{\gamma'}^{\theta}},
\label{ddbux}
\end{eqnarray}
where $\varepsilon$ is a three-body breakup energy measured from the ${^7\mathrm{Be}}+p+p$ threshold, 
whereas $\Omega$ is the solid angle of the c.m. of the three particles.
The complex-scaled $T_{\gamma}$ are given by
\begin{align}
T_{\gamma'}^{\theta}
&=
\sum_\gamma T_\gamma^*
\left\langle \Phi_{\gamma}
\Big|
C^{-1}(\theta)
\Big|
\varphi_{\gamma'}^{\theta}
\right\rangle,
\label{csm-T1}\\
\tilde{T}_{\gamma'}^{\theta}
&=
\sum_\gamma \left\langle \tilde{\varphi}_{\gamma'}^{\theta}
\Big|
C(\theta)
\Big|
\Phi_{\gamma}
\right\rangle
T_\gamma
\label{csm-T2}
\end{align}
with $C(\theta)$ and $C^{-1}(\theta)$ being the scaling-transformation
operator and its inverse, respectively~\cite{Mat10}, and are defined as
\begin{eqnarray}
 \langle r_c,y_c|C(\theta)|f\rangle
  &=&e^{3i\theta}f(r_ce^{i\theta},y_ce^{i\theta}),\\
 \langle f|C^{-1}(\theta)|r_c,y_c\rangle
  &=&\left[e^{-3i\theta}f(r_ce^{-i\theta},y_ce^{-i\theta})\right]^{*},
\end{eqnarray}
where function $f$ is $\Phi_\gamma$. As shown in Eq.~(\ref{ddbux}), the breakup energy spectrum is given by an incoherent sum of the contributions from the eigenstates of
$h^{\theta}$. This property is crucial to clarify the role of
a resonance in describing breakup observables.

\section {Results and discussions} 
\subsection{Binary and ternary potentials} \label{3I}
Analysis of the $^8$B (${^7\mathrm{Be}}+p$) subsystem is imperative in studying the structure of $^9$C as the three-body system. 
The measured spectrum of $^8$B consists of only one weakly-bound state with $I^\pi=2^+$ and the one-proton separation energy $S_p$ is $0.137$\,MeV.
Thus, in an independent-particle shell model picture, $^8$B is usually 
described as a valence proton in the $0p_{3/2}$ orbit loosely bound to the $^7$Be core (ground state with $I^\pi=3/2^-$)~\cite{Baye94}. 

In the present study, for simplicity, we assume that the valence proton in the $0p_{3/2}$ orbit is coupled to the inert and spinless $^7$Be core. Note that a similar assumption has been successfully followed in other three-body calculations, for instance, the $^9$Li$+n$ and $^{27}$F$+n$ subsystems for the description of $^{11}$Li~\cite{Mat19} and $^{29}$F~\cite{Casal20}, respectively. 
The numerical detail of the GEM relevant to the calculation of the binary systems is relegated to Sec.~\ref{3II} (see Table~\ref{Tbasis}). 

As regards the interaction between the $^7$Be core and a valence proton ($V_{cp}$), 
we adopt the same nuclear potentials used in Refs.~\cite{Esb96,Gold07}, including the 
central and spin-orbit components. For the Coulomb potential, we take 
$V_{\mathrm{C}}^{cp}(x)=4(e^2/x)~{\rm erf}(sx)$ with $s=0.65$\,fm$^{-1}$, 
where $x$ stands for the relative $^7$Be-$p$ distance.
With this interaction, we obtain the $0p_{3/2}$-ground state with $S_p=0.137$\,MeV.
It should be noted that in the absence of the core spin, all $0^+$, $1^+$, $2^+$, and $3^+$ states (corresponding to coupling of the core spin $3/2$ with valence-proton spin $3/2$) are degenerate with the ground state. 
Thus, the low-lying $1^+$, $0^+$, and $3^+$ resonances~\cite{Mitchell13,Gold98} above the ${^7\mathrm{Be}}+p$ threshold are not reproduced by this model.

Interestingly, with the same interaction, we obtain a two-body resonance of $^8$B in the $0p_{1/2}$ state with resonant energy (decay width) of $2.270$\,MeV ($1.450$\,MeV). Similarly, in the absence of core spin the possible degenerate states corresponding to this state are $1^+$ and $2^+$. 
However, this two-body resonance is found to play no roles in the breakup calculations, though a feature of this resonance can be seen inside a three-body system (see discussion in Sec.~\ref{3III}). In other partial waves, resonances do not exist. 

The second binary interaction used in the internal Hamiltonian $h$ of $^9$C is the $p$-$p$ interaction ($V_{pp}$), 
for which we adopt the Minnesota (MN) interaction~\cite{Tho77} and a Coulomb potential between two protons. 
For the MN interaction, the exchange-mixture parameter $u$ is set equal to $0.95$ and only 
its central part is included, 
while, for the Coulomb part, we take $V_{\mathrm{C}}^{pp}(x)=e^2/x$, with the relative $p$-$p$ distance $x$. 

In addition to $V_{cp}$ and $V_{pp}$, it is customary to 
introduce a ternary potential ($V_{cpp}$) to account for possible 
effects that are not explicitly included in the above-mentioned three-body description~\cite{Gal05, Diego10, Casal13, Mat14}. 
We take volume-type 3BF~\cite{Mat14} given by a product of Gaussian functions for the two Jacobi coordinates; $V_0$~$\exp(-\alpha r_1^2-\alpha y_1^2)$.
We set the range parameter $\alpha=0.0357$\,fm$^{-2}$ and the strength $V_0$ is determined to optimize 
the ground state energy and low-lying resonance energies; see Sec.~\ref{3II}. This flexibility of 
the adopted three-body model is preferable because the reproduction of threshold energies is crucial for describing reaction properties of $^9$C. 

\subsection{Three-body ground state of ${^9\mathrm{C}}$}
\label{3II}
In the present three-body model with the inert-spinless core approximation, our ground state is represented by coupling the valence protons to $0^+$ and the continuum-excited states correspond to the $0^+$, $1^-$, and $2^+$ states. 

In the GEM, the ground state of the ${^7\mathrm{Be}}+p+p$ system is described by 
a superposition of three channels, each of which is specified by a certain set of the Jacobi coordinates, $(\bm{r}_c,\bm{y}_c)$.  
In the channel $c$, the radial parts of the internal wave functions 
involving $\bm{r}_c$ and $\bm{y}_c$ are expanded by a finite number of Gaussian basis functions as
\begin{align}
 \varphi_{i\ell}(\bm{r}_c)=&
  r_c^\ell e^{-(r_c/\bar{r}_i)^2}Y_\ell(\Omega_{r_c}),
  \label{GEM1}\\
  \varphi_{j\lambda}(\bm{y}_c)=&y_c^\lambda e^{-(y_c/\bar{y}_j)^2}
  Y_\lambda(\Omega_{y_c}).
  \label{GEM2}
\end{align}
Here, $\ell$ ($\lambda$) and $\Omega_{r_c}$ ($\Omega_{y_c}$) are the orbital-angular momentum and solid angle, respectively, associated with the coordinate $\bm{r}_c$ ($\bm{y}_c$), 
and the range parameters are given by the geometric progression:
\begin{align}
 \bar{r}_i&=(\bar{r}_{\rm max}/\bar{r}_{\rm 1})^
  {(i-1)/i_{\rm max}},\label{para1}\\
 \bar{y}_j&=(\bar{y}_{\rm max}/\bar{y}_{\rm 1})^
  {(j-1)/j_{\rm max}}.\label{para2}
\end{align}
To be precise, the parameters depend on $c$, but we omitted the dependence 
in Eqs. \eqref{para1} and \eqref{para2} for simplicity; 
see Ref.~\cite{Mat04} for the details of 
the diagonalization and the definition of the Jacobi coordinates. 

\begin{table}[!t]
 \caption{Parameters of the GEM for the $0^{+}$,~$1^{-}$,~and~$2^{+}$ states.} 
 \begin{tabular}{c|ccccccc}
  \toprule\\[-1.5ex]
  Set&$c$&$j_{\rm max}$&
  $\bar{y}_1$ (fm)&$\bar{y}_{\rm max}$ (fm)&$i_{\rm max}$
  &$\bar{r}_1$ (fm)&$\bar{r}_{\rm max}$ (fm)
  \\ \hline\hline
  I &3&13&   0.1&       15.0& 13&   0.5&  15.0   \\
  &1, 2&13&   0.5&      15.0&13&    0.5&  15.0 \\ \hline
 II &3 &25&   0.1&       50.0& 25&   0.5&  50.0   \\
  &1, 2&25 &  0.5&       50.0& 25&   0.5&  50.0 \\ [1ex]
\botrule
 \end{tabular}
\label{Tbasis}
\end{table}

For the evaluation of eigenstates of $h$ and $h^{\theta}$, we adopt the parameter
set I for $h$  and set II for $h^{\theta}$ listed in Table~\ref{Tbasis}. 
\begin{table}[h!]
\caption{The ground-state energy of $^9$C and its interaction dependence.}
\centering
\begin{tabular}{lc}
\toprule\\[-1.5ex]
Interaction  & Ground state energy  \\ 
& $S_{2p}$(MeV) \\ [1ex]
\colrule\\[-1.5ex]
%
$V_{pc}+V_{pc}$ (without $V_{pp}$) & $-0.254$\\
$V_{pc}+V_{pc}+V_{pp}$& $-3.070$\\
$V_{pc}+V_{pc}+V_{pp}+V_{cpp}$&$-1.437$\\ [1ex]
\botrule
\end{tabular}
\label{Tint}
\end{table}

Using the binary ($V_{cp}$ and $V_{pp}$) and ternary ($V_{cpp}$) interactions, described in Sec.~\ref{3I}, 
we compute the three-body ground state of the $^9$C nucleus with $I^\pi=0^+$. 
In Table~\ref{Tint}, the interaction dependence of the ground-state energy is displayed.
It is clear that the role of $V_{pp}$ is significant in binding the two protons in $^9$C and a repulsive three-body force is needed to obtain the experimental value of the two-proton separation energy, $S_{2p}=1.437$\,MeV. The need of repulsive three-body force is not so rare; the same situation has been encountered in the three-body calculations of $^{12}$C \cite{Kurokawa07, Ohtsubo13}.
The 3BF strength optimized for the ground state is obtained as $V_0=3.172$\,MeV to reproduce the experimental $S_{2p}$. Note that, while this value of $V_0$ is employed also for the $1^-$ states,
$V_0$ for the $2^+$ states are optimized independently, as described in Sec.~\ref{3III}. 

\subsection{Structure of $0^{+}$,~$1^{-}$,~and~$2^{+}$ continua}
\label{3III}
\begin{figure}[!t]
\includegraphics[scale=0.35]{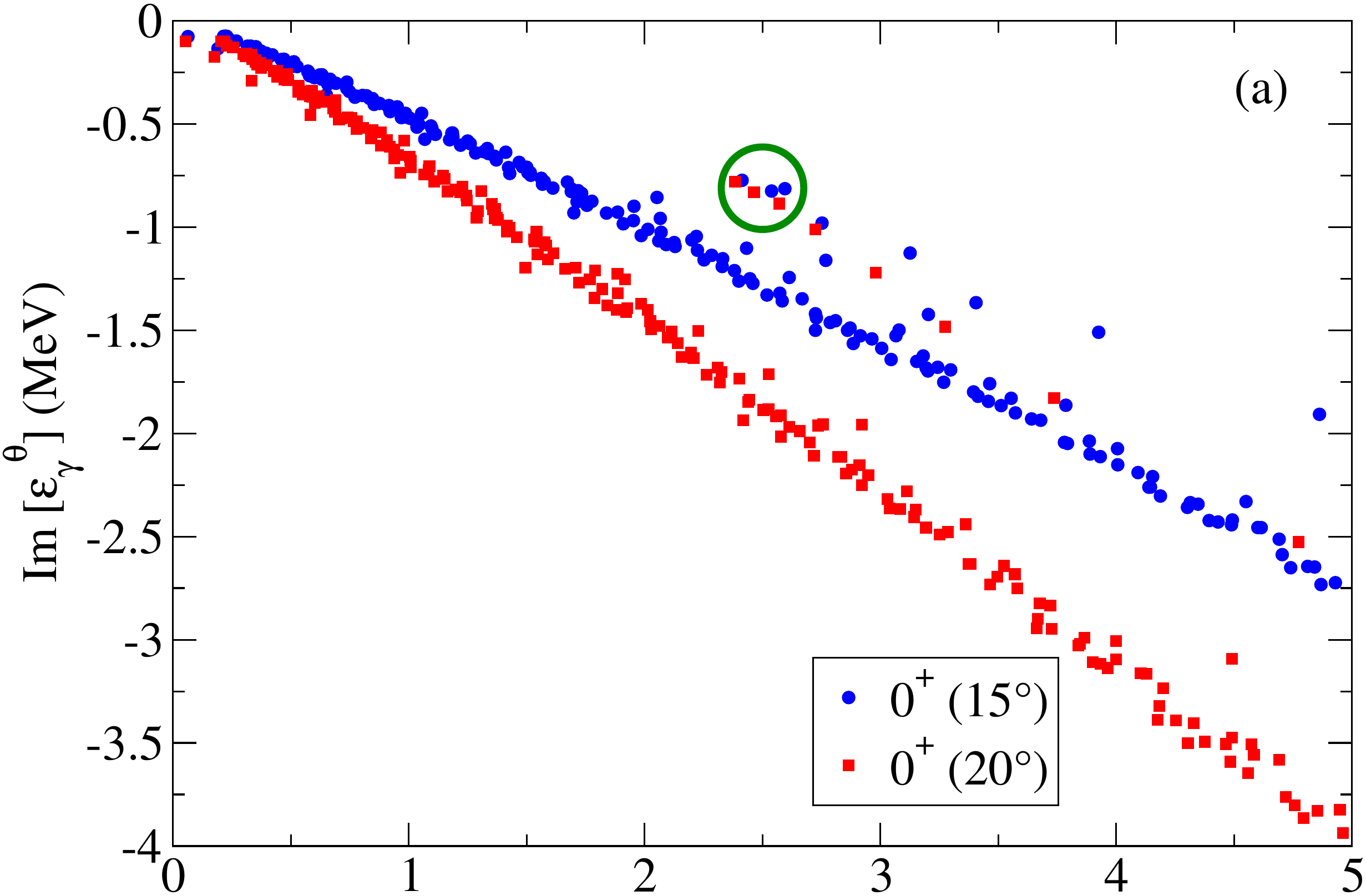}
\includegraphics[scale=0.35]{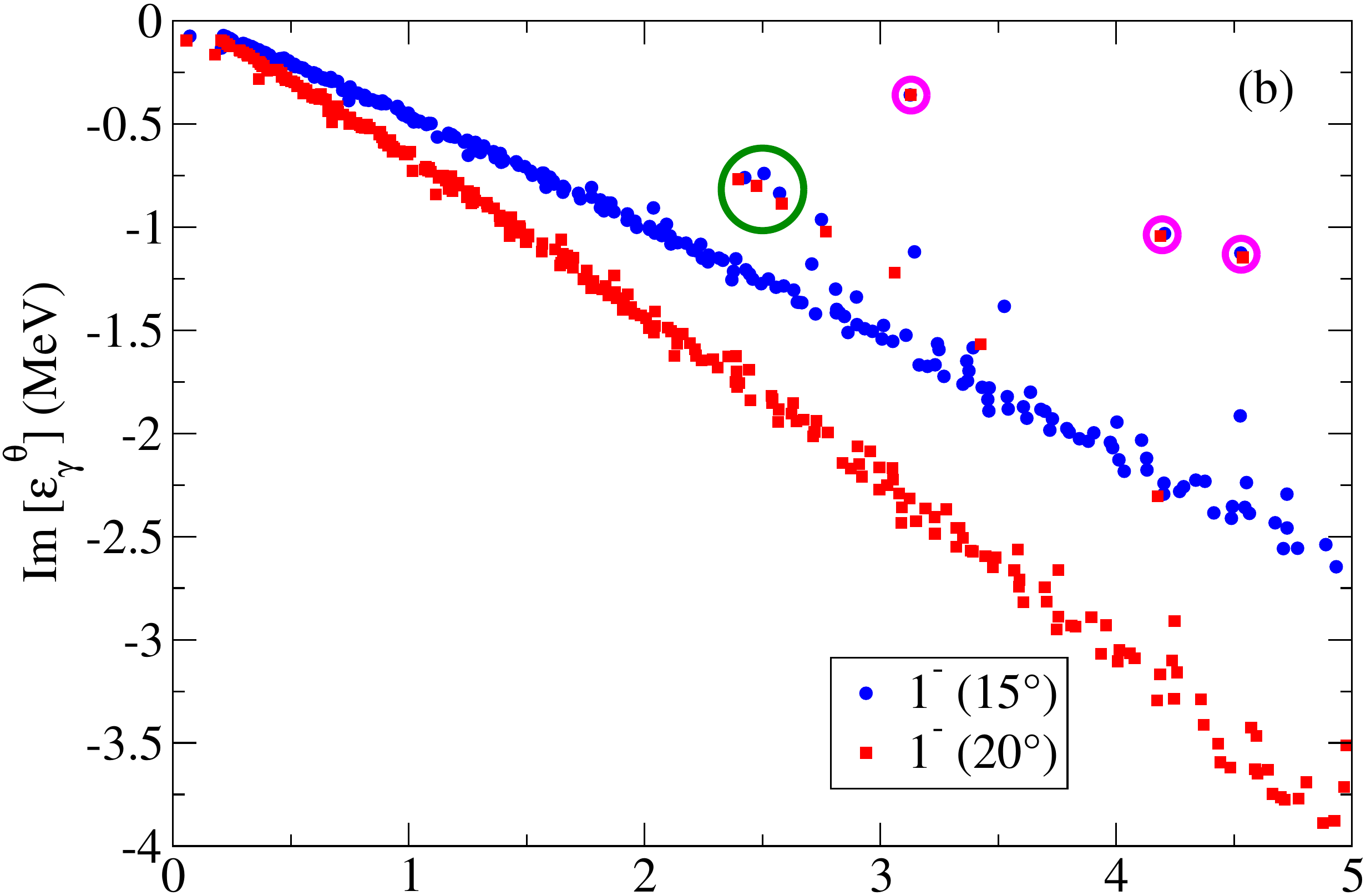}
\includegraphics[scale=0.35]{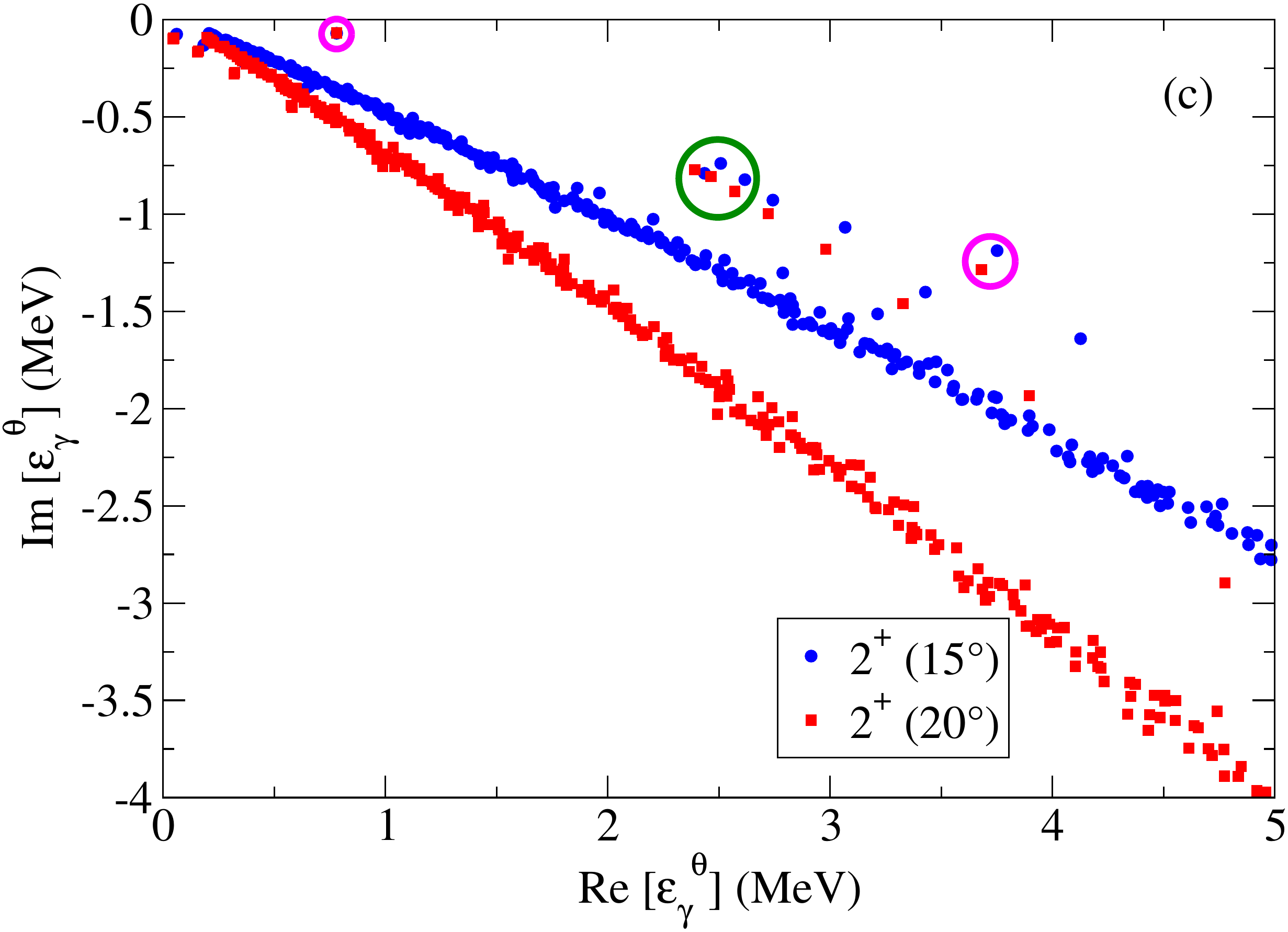}
\caption{
Complex eigenenergies of $^9$C for (a) $I^{\pi}=0^{+}$, (b) $I^{\pi}=1^{-}$, and (c) $I^{\pi}=2^{+}$ states with the scaling angle $\theta=15^{\circ}$ (blue circles) and $20^{\circ}$ (red squares). For details see text.}
\label{fig3}
\end{figure}
After fixing the $0^+$-ground state of $^9$C, 
the eigen energies of resonant and nonresonant continuum states 
for $I^{\pi}=0^{+}$, $1^{-}$, and $2^{+}$, which are respectively displayed in Figs.~\ref{fig3}(a), (b), and (c), 
are computed by using the GEM combined with the CSM.
In Fig.~\ref{fig3}, the blue circles and red squares represent the eigenenergies with the scaling angle $\theta=15^{\circ}$ and $20^{\circ}$, respectively and $\theta=20^{\circ}$ is the converged value 
with respect to the position of all the resonances. 
In the converged model space, the total number of pseudostates below $\varepsilon_{\gamma}=7$\,MeV are $40$, $50$, and $59$ for 
$I^{\pi}=0^{+}$, $1^{-}$, and $2^{+}$, respectively. These states are included in the 
calculation of the breakup cross sections shown in Sec.~\ref{3IV}.

Three two-body resonances (encircled by green circles in Fig.~\ref{fig3}) are found in the $0^+$, $1^-$, and $2^+$ states, 
which are found to be unaffected by the inclusion of the 3BF. 
They correspond to the same $^8$B resonant state in $p_{1/2}$ at around $~2.3$\,MeV. 
Depending on the angular momentum of the other proton in nonresonant states, 
the total spin-parity can be $0^+$ ($p_{1/2}$), $1^-$ ($s_{1/2}$ or $d_{3/2}$), and $2^+$ ($p_{3/2}$ or $f_{5/2}$).

\begin{figure}[!b]
 \includegraphics[scale=0.35]{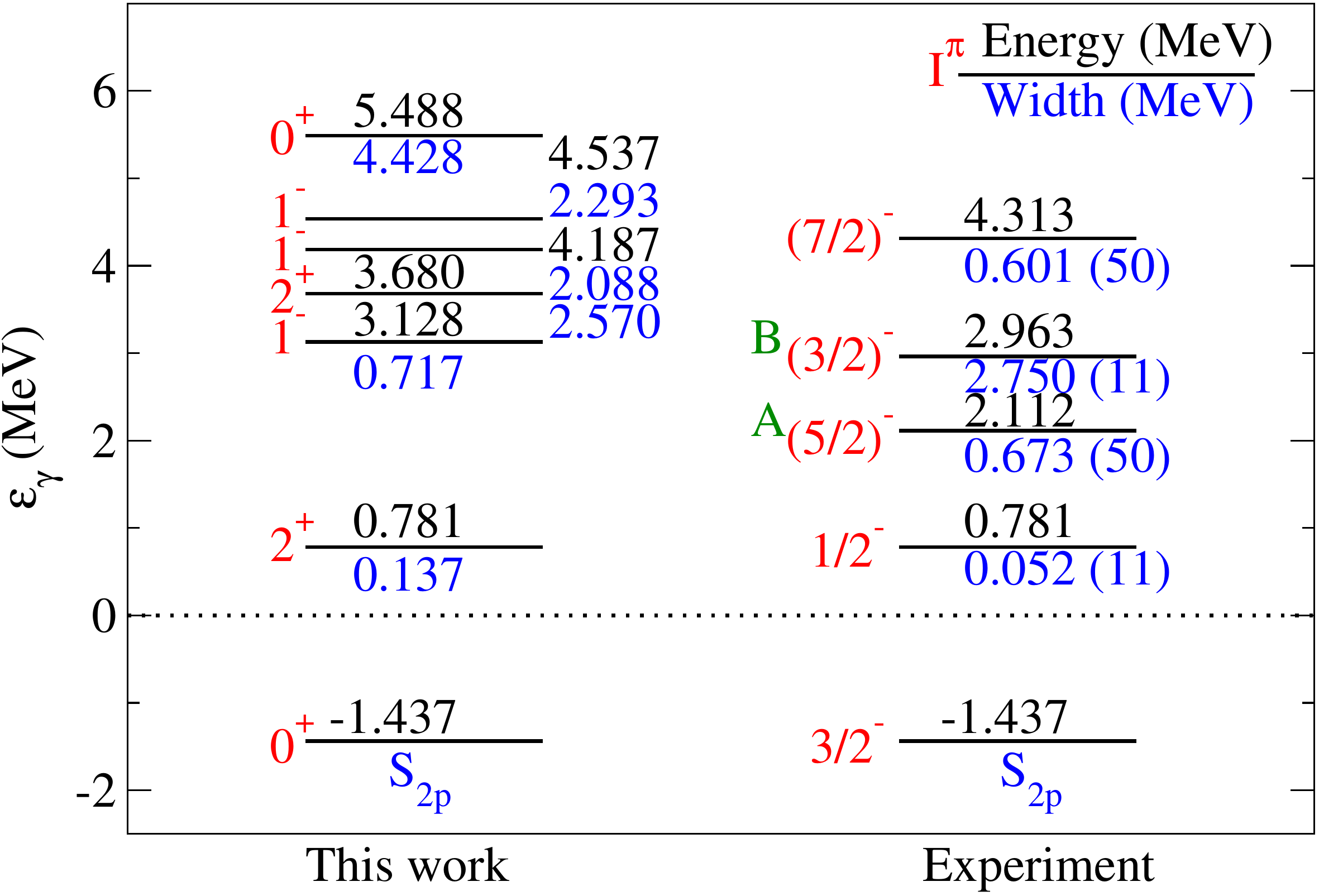}
 \caption{ The comparison of our three-body results with the available experimental data adopted from Ref.~\cite{Brown17}.
 $^A$In Ref.~\cite{Rogachev07}, a $5/2^-$ state at $2.163$ ($2$)\,MeV, with a width of $1.4$ ($5$)\,MeV is reported and $^B$in Ref.~\cite{Hooker19}, a broad positive-parity state $5/2^+$ at $2.863$\,MeV, is reported. The experimental uncertainties are in units of keV.}
\label{fig}
\end{figure}
A broad three-body resonance is found in the $0^+$ continuum with the resonant energy $5.488$\,MeV and the decay width $4.428$\,MeV (not shown in Fig.~\ref{fig3}(a)), with the inclusion of three-body force. In the region of $\epsilon_\gamma > 3$\,MeV, three three-body resonances (encircled by pink circles in Fig.~\ref{fig3}(b)) are obtained in the $1^-$ continuum.
Their resonant energies (decay widths) in the units of MeV are $3.129$ ($0.719$), $4.187$ ($2.088$), and $4.537$ ($2.293$). 
It is found that the position of the $1^-$ resonances are weakly dependant on the strength of the 3BF. 
This is because these resonances have an extended structure (\textit{i.e.,} three constituents are not in close proximity of each other). On the other hand, the above-mentioned $0^+$ resonance has a compact form (\textit{i.e.,} three constituents stay close to each other), which shifts to higher energy by including the 3BF. It should be noted that, for the $0^+$- and $1^-$-continuum states, we employ the same 3BF strength ($V_0=3.172$\,MeV) determined for the ground state.  

The narrow low-lying and relatively high-lying wider three-body resonances (encircled by pink circles in Fig.~\ref{fig3}(c)) are obtained in the $2^+$ continuum. Their resonant energies (decay widths) in the units of MeV are $0.781$ ($0.137$) and $3.680$ ($2.088$). 
For the $2^+$-continuum states, the 3BF strength is optimized as $V_0=5.320$\,MeV to obtain the low-lying experimental resonance energy, and this strength is different from that used for $0^+$ and $1^-$ states. 
Without the 3BF, the low-lying $2^+$ narrow resonance becomes a bound state and the high-lying $2^+$ wider resonance moves to lower energy ($\epsilon_\gamma = 1.44$\,MeV). 
Similar to the $0^+$ states, the effect of the 3BF is strong and clearly visible for the $2^+$ states, 
and this is due to the compact structures of the optimized bound states. 

Figure~\ref{fig} shows the comparison of our results with the available experimental data~\cite{Brown17}. 
The energies (shown in black color) and widths (shown in blue color) are given above and below the lines, respectively, in units of MeV. As 
mentioned, we have tuned the 3BF so as to reproduce the energies of the ground state and the first excited state. The width of the later 
thus found to be consistent with the experimental value. As for relatively high-lying states, the preceding experimental study \cite{Brown17} suggested $5/2^-$, $3/2^-$, and $7/2^-$ states. 
The first one has been confirmed in Ref.~\cite{Rogachev07} but with a much larger width. Furthermore, recently, a $5/2^+$ state is reported at $2.863$\,MeV \cite{Hooker19}. Thus, these states are still under discussion. Our three-body model calculation with a spinless core approximation suggest relatively high-lying sole $0^+$, one $2^+$, and three $1^-$ resonances, whose parity will be negative, negative and positive, respectively, when we consider the spin-parity of $^7$Be. A further investigation of these possible resonance states through a breakup cross section will be interesting and important.

\subsection{${^9\mathrm{C}}+{^{208}\mathrm{Pb}}$ elastic breakup}
\label{3IV}
Next, we perform the four-body CDCC calculation for the $^9$C scattering on the $^{208}$Pb target at $65$\,MeV/nucleon and $160$\,MeV/nucleon, using the wave functions of the bound and discretized-continuum states of $^9$C described in Secs.~\ref{3II} and \ref{3III}. 
In this analysis, the convergence of the calculated elastic breakup cross section has been achieved within about $3.5\%$ fluctuation with respect to the GEM parameters tabulated in Table~\ref{Tbasis}. We truncated the eigenenergy of the pseudostates at $7$\,MeV. 

The distorting nucleon-nucleus ($p$-$^{208}{\rm Pb}$) and nucleus-nucleus ($^7{\rm Be}$-$^{208}{\rm Pb}$) potentials are evaluated by a microscopic folding model. The Melbourne nucleon-nucleon $g$ matrix~\cite{Amo00} and the Hartree-Fock wave functions of $^7{\rm Be}$ and $^{208}{\rm Pb}$ based on the Gogny D1S force~\cite{DG80,Ber91} are adopted. This microscopic approach has successfully been applied to several reaction systems~\cite{Yah12,Min12,Sum12}.

\begin{figure}[!t]
 \includegraphics[scale=0.35]{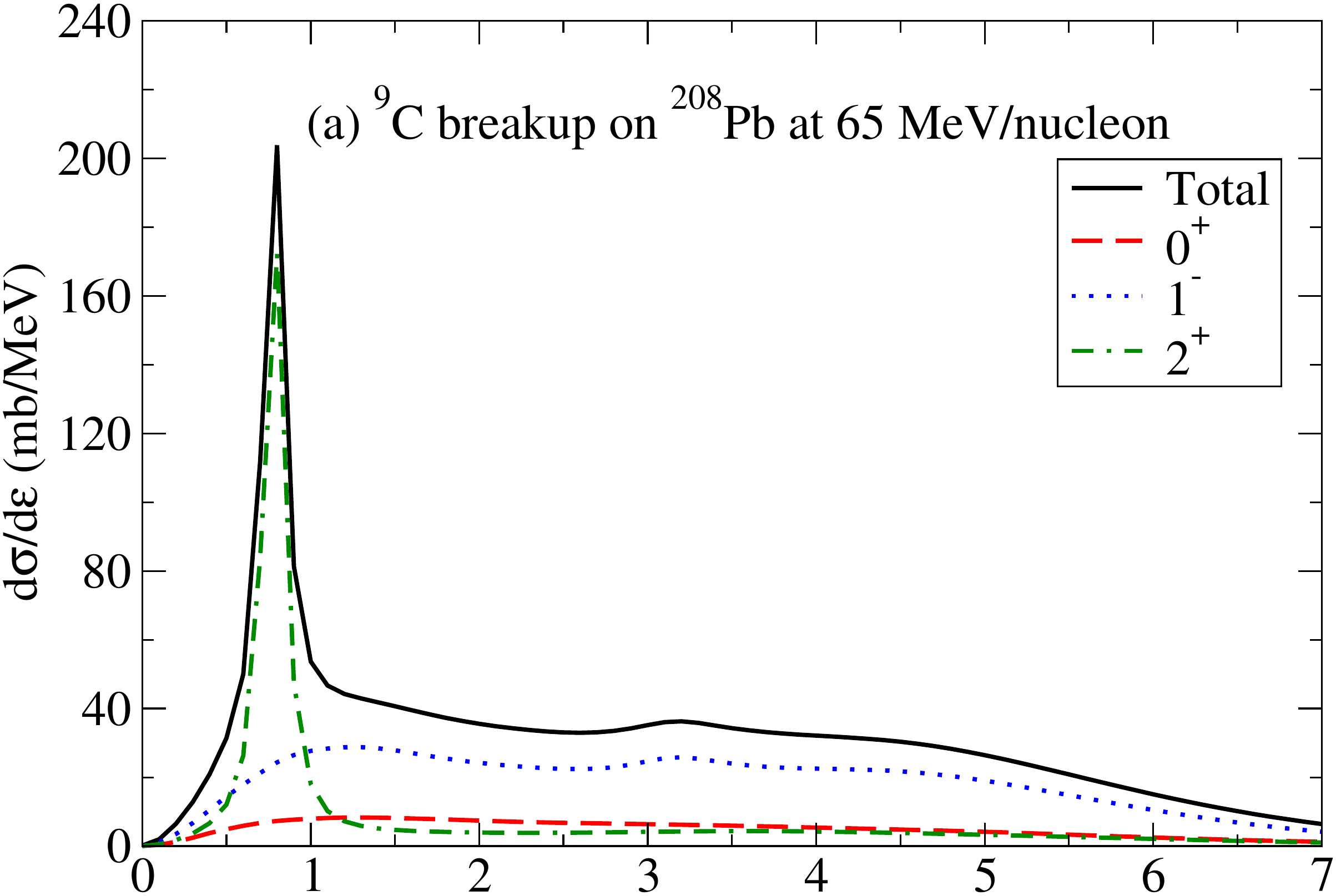}
 \includegraphics[scale=0.35]{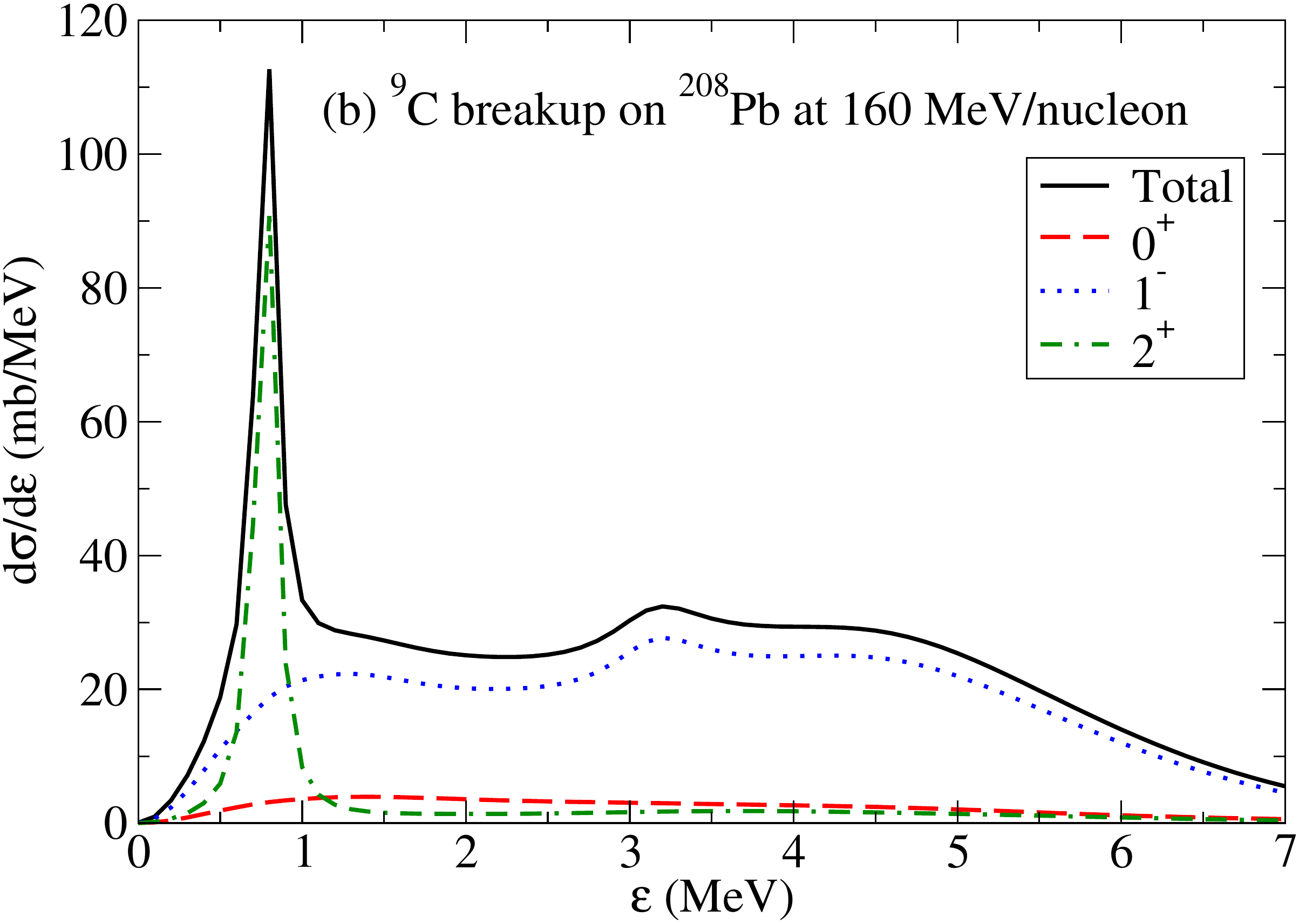}
 \caption{Energy spectra of the breakup cross section 
 of $^{9}$C on $^{208}$Pb at 
(a) $65$\,MeV/nucleon and (b) $160$\,MeV/nucleon. The solid, 
dashed, dotted, and dash-dotted lines correspond to the total, $0^+$, 
$1^-$, and $2^+$ breakup contributions, respectively.}
 \label{fig4}
\end{figure}

We investigate the role of the $0^+$, $1^-$, and $2^+$ resonant and nonresonant-continuum states on the cross section. Figures~\ref{fig4}(a) and~(b) show the elastic breakup cross section at $65$ and $160$\,MeV/nucleon, respectively, 
as a function of the smoothed three-body breakup energy $\varepsilon$.
The curves are obtained by integrating the double differential cross section, Eq.~\eqref{ddbux}, over $\Omega$.
We can decompose the cross section into each contribution of the spin-parity state of $^9$C; 
the solid, dashed, dotted, and dash-dotted lines denote the 
the total, $I^{\pi}=0^{+}$, $1^{-}$, and $2^{+}$ contributions, respectively. 

Now, let us focus on the case of the lower incident energy $65$\,MeV/nucleon in Fig.~\ref{fig4}(a). 
It is well known that, in the breakup reactions caused by a heavy target, the $E1$ transitions are expected to be a dominant contribution. In our model, because the spin-parity of the ground state of $^9$C is $0^+$ the large dominance of the $1^-$ contribution is obtained as about $58.4\%$ of the total breakup cross section. 
The shape of the breakup cross section is guided by a narrow peak around $0.781$\,MeV in the lower energy region, namely $\varepsilon \leq$ $2$\,MeV. 
This peak originates from the narrow $2^+$ resonance, 
and its contribution to the total breakup cross section is around $19.0\%$ (whereas full $2^+$ contribution is about $26.0\%$ to the total breakup cross section). In the higher energy region, say, $\varepsilon >$ $2$\,MeV, the shape of the total breakup cross section is mainly determined by the $1^-$ resonance at $3.129$\,MeV. Thus, we find that the $1^{-}$-cross section is dominated by the low-lying nonresonant continuum and high-lying $1^-$ resonance state at $3.129$\,MeV. 
Furthermore, the $0^{+}$-cross section is not negligible (about $15.6\%$ of the total breakup cross section) and comes from nonresonant continuum because no low-lying three-body resonances exist for $I^{\pi}=0^{+}$. 

The same features can be seen at the higher incident energy $160$\,MeV/nucleon, as shown in Fig.~\ref{fig4}(b). 
Quantitatively, the magnitude of breakup cross sections are smaller with respect to the lower 
incident energy case. However, the peak in $1^-$ breakup at $3.129$\,MeV is more pronounced in the $160$ MeV/nucleon case.

In summary, the first $2^{+}$ state with a small decay width is well understood as the first excited state, and its contribution to the cross sections shows up as a sharp peak. The high-lying second $2^+$ and three $1^{-}$ resonant states contribute to the cross section in the higher breakup-energy region of the energy spectrum as a bump-shape structure. Whereas, the high-lying $0^+$ resonant state contributes negligibly small to the total cross sections in the energy region, viz., $\varepsilon \geq$ $5$\,MeV and hence does not show any structural feature in the breakup cross sections. Thus, we conclude that, in order to clarify the properties of the resonant states via breakup observables, it is required an accurate analysis of treating not only resonant contributions but also nonresonant ones in the energy spectrum. 

As a future work we need to investigate the role of the spin of the core nucleus $^7$Be. 
We will follow the prescription proposed in Ref.~\cite{Ogata06},
in which the spin of the  $^7$Be core is simplified in peripheral breakup reactions 
to extract the astrophysical $S$-factor of ${^7\mathrm{Be}}(p,\gamma){^8\mathrm{B}}$. 
Moreover, it is necessary to disentangle the ${^8\mathrm{B (g.s.)}}+p$ channel 
from the discretized continuum channels in the calculation of the breakup cross section.
To this end, a practical method proposed in Ref.~\cite{watanabe2021practical} is expected to provide 
useful information.
\section{Conclusions}
We have investigated the ground and low-lying continuum states 
of the $^9$C nucleus by means of the GEM combined with the CSM. 
The role of these low-lying resonances is studied in the elastic breakup of 
$^9$C on the $^{208}$Pb target at $65$ and $160$\,MeV/nucleon, using the CDCC based on the ${^7\mathrm{Be}}+p+p+{^{208}\mathrm{Pb}}$ four-body model.

As a result of the CSM with disregarding the spin of ${^7\mathrm{Be}}$, 
we have obtained the resonant energy and decay width of the first $2^{+}$ state,
which is consistent with the available experimental information. 
We have also found high-lying $0^+$, second $2^+$, and additional three  $1^-$ resonant states above the first $2^+$ state.

In the analysis of $^9$C breakup on $^{208}$Pb, we have calculated the elastic breakup cross section to  investigate the contribution of the low-lying resonances to the energy spectrum. 
We have shown that the contribution of nonresonant continuum states to the breakup cross section is also important. 
In conclusion, an accurate analysis of treating both resonances and nonresonant continuum states is highly required to clarify the properties of the energy spectrum of breakup cross sections.
In future, it will be important to extract astrophysical information from the present result.
Moreover, it would be interesting to calculate two-proton removal cross sections with the present four-body model and compare it with the new upcoming experimental data~\cite{Chilug19,Chilug20}.

\section*{Acknowledgements} 
The authors would like to thank N. Timofeyuk for her important remarks on the future perspectives of this work. The computation was carried out using the computer facilities at the Research Center for Nuclear Physics, Osaka University. 
This work is supported in part by Grant-in-Aid for Scientific Research (No.~JP18K03650) from Japan Society for the Promotion of Science (JSPS). 
\bibliography{ref}
\end{document}